\documentclass{article_saj}
\usepackage{graphicx,saj,multicol,subeqnarray}
\usepackage{natbib}
\usepackage{float}
\usepackage{footmisc}
\usepackage{xcolor}
\usepackage{widetext}
\usepackage{url}
\usepackage{bm}
\usepackage{tikz} 
\usepackage{pifont} 
\usepackage{amsfonts}
\usepackage{amssymb}
\usepackage{graphicx}
\usepackage{amsmath,upgreek}
\usepackage{titlesec}
\titlelabel{\thetitle.\quad}
\definecolor{xlinkcolor}{cmyk}{1,0.6,0,0}
\usepackage[bookmarks=false,         
     pdfnewwindow=true,      
     colorlinks=true,    
     linkcolor=xlinkcolor,     
     citecolor=xlinkcolor,     
     filecolor=xlinkcolor,  
     urlcolor=xlinkcolor,      
final=true
]{hyperref}

\newcommand{\ergcm}[1]{erg\,cm$^{-2}$\,s$^{-1}$}

\def\HI{\hbox{H{\sc i}}}
\def\HII{\hbox{H{\sc ii}}}

\newcommand{\SII}{[S\,{\sc ii}]}

\newcommand{\NII}{[N\,{\sc ii}]}
\newcommand{\Halpha}{H${\alpha}$}

\def\papertype{\ \hfill\ Editorial}

\def\udc{52}
\renewcommand{\thefootnote}{\fnsymbol{footnote}}

\begin{document}
\parindent=.5cm
\baselineskip=3.8truemm
\columnsep=.5truecm
\newenvironment{lefteqnarray}{\arraycolsep=0pt\begin{eqnarray}}
{\end{eqnarray}\protect\aftergroup\ignorespaces}
\newenvironment{lefteqnarray*}{\arraycolsep=0pt\begin{eqnarray*}}
{\end{eqnarray*}\protect\aftergroup\ignorespaces}
\newenvironment{leftsubeqnarray}{\arraycolsep=0pt\begin{subeqnarray}}
{\end{subeqnarray}\protect\aftergroup\ignorespaces}
%


\markboth{\eightrm Radio continuum Study of the Large Magellanic Cloud Supernova Remnant Honeycomb Nebula} 
{\eightrm R. Z. E. Alsaberi {\lowercase{\eightit{et al.}}}}

\begin{strip}

{\ }

\vskip-1cm

\publ

\type

{\ }


\title{Radio Continuum Study of the Large Magellanic Cloud Supernova Remnant Honeycomb Nebula}


\authors{R. Z. E. Alsaberi$^{1}$, M. D. Filipovi\'c$^{1}$, H. Sano$^{2,3}$, P. Kavanagh$^{4}$, P. Janas$^{4}$, J. L. Payne$^{1}$, D. Uro\v{s}evi\'c$^{5}$}

\vskip3mm


\address{$^1$Western Sydney University, Locked Bag 1797, Penrith South DC, NSW 1797, Australia}

\Email{19158264@student.westernsydney.edu.au, m.filipovic@westernsydney.edu.au, astronomer@icloud.com}

\address{$^2$Faculty of Engineering, Gifu University, 1-1 Yanagido, Gifu 501-1193, Japan}
\address{$^3$Center for Space Research and Utilization Promotion (c-SRUP), Gifu University, 1-1 Yanagido, Gifu 501-1193, Japan}
\Email{sano.hidetoshi.w4@f.gifu-u.ac.jp}

\address{$^4$Department of Experimental Physics, Maynooth University, Maynooth, Co.\ Kildare, Ireland}
\Email{patrick.kavanagh@mu.ie, pawel.janas.2020@mumail.ie}

\address{$^5$Department of Astronomy, Faculty of Mathematics, University of Belgrade, Studentski trg 16, 11000 Belgrade, Serbia}
\Email{dejan.urosevic@matf.bg.ac.rs}



\dates{May 18, 2020}{June 1, 2020}


\summary{We present the first and deepest Australia Telescope Compact Array radio continuum images of the Honeycomb Nebula at 2000 and 5500\,MHz solely from archival data. The resolutions of these images are $3.6\times2.8$\,arcsec$^2$ and $1.3\times1.2$\,arcsec$^2$ at 2000 and 5500\,MHz. We find an average radio spectral index for the remnant of $-0.76\pm0.07$. Polarisation maps at 5500\,MHz reveal an average fractional polarisation of $25\pm5$\% with a maximum value of $95\pm16$. We estimate the equipartition field for Honeycomb Nebula of $48\pm5$\,$\mu$G, with an estimated minimum energy of $E_{\rm min}=3\times10^{49}$\,erg. The estimated surface brightness, $\Sigma_{\rm 1\,GHz}$, is $30\times10^{-20}$\,W\,m$^{-2}$\,Hz$^{-1}$\,sr$^{-1}$; applying the $\Sigma$-D relation suggests this supernova remnant is expanding into a low-density environment. Finally, using \HI\ data, we can support the idea that the Honeycomb Nebula exploded inside a low-density wind cavity. \textbf{We suggest that this remnant is likely to be between late free expansion stage and early Sedov phase of evolution and expanding into a low-density medium.} }



\end{strip}

\tenrm


\section{INTRODUCTION}
\renewcommand{\thefootnote}{\arabic{footnote}}
\indent

Supernova Remnants (SNRs) play an essential role in  galaxies, enriching the Interstellar Medium (ISM)  producing a significant impact on their  structure and physical properties. However, the study of SNRs within our own Galaxy is not ideal because of a high level of dust and gas absorption, compounding the difficulties of achieving accurate distance measurements. Instead, we look to nearby galaxies, such as the Large Magellanic Cloud (LMC), located at a known distance of $\sim$50\,kpc \citep{mt,2019Natur.567..200P}. This  allows observers to assume  all objects within the galaxy are located at the same distance, making physical measurements, including the extent of physical size, more reliable. Additionally, the LMC is  approximately face-on in orientation \citep[having an inclination angle of $\sim$35$^{\circ}$,][]{2001AJ....122.1807V} and is near enough to allow deep, high-resolution (spatial and spectral)  multi-frequency SNR observations \citep{2016A&A...585A.162M,2017ApJS..230....2B,2021MNRAS.500.2336Y,2023MNRAS.518.2574B}.

The Honeycomb Nebula (MC\,SNRJ0535--6918) is one of the most peculiar SNRs in the LMC -- just a few arcminutes south-east of the well-known SN\,1987A. It was discovered by \cite{1992Msngr..69...34W} using the {\it ROSAT} survey. \cite{1993MNRAS.263L...6M} presented a kinematic study of the Honeycomb Nebula and  concluded that all the kinematic features can be generated as one or more young SNRs colliding in a dense environment. \cite{1995AJ....109.1729C} used a combination of archival and new {\it ROSAT} X-ray data and  Australia Telescope Compact Array (ATCA) with Molonglo Observatory Synthesis Telescope (MOST) radio data to find both bright X-ray and nonthermal radio emission. Moreover, they reported an enhanced \SII/\Halpha\ ratio and a steep radio spectral index of --1.2. 
\cite{1999A&A...345..943R} used optical observations obtained from the \textit{Manchester Echelle} spectrometer to produce new kinematical and density data for the Honeycomb Nebula.  \cite{2010MNRAS.408.1249M} also used optical data in a spectral analysis of forbidden line ratios to infer the  nebula is most likely a SNR. 

\cite{2010MNRAS.408.1249M}  found  the Honeycomb Nebula is more extended in soft X-ray emission and could be a small section on the edge of a giant LMC shell. They suggest a secondary supernova explosion on this edge may be responsible for the creation of the nebula.

In this paper we present a radio continuum study of the Honeycomb Nebula using ATCA archival data to produce high-fidelity images and we compare them with other multiwavelength data (X-ray and optical). In Section~\ref{data} we present the data used and their reduction details. Section~\ref{res} explores further analysis  and discussion including radio morphology, polarisation, spectral index, surface brightness and luminosity. Finally, in Section~\ref{con}, we conclude with a summary of our findings.

\section{DATA} 
\label{data}

\subsection{Radio continuum observations}
\label{atca}

\indent

The Honeycomb Nebula has been serendipity observed with ATCA, due to its location in the field of view of SN\,1987A \citep{1995AJ....109.1729C}. Archival\footnote{Australia Telescope Online Archive (ATOA), hosted by the Australia Telescope National Facility (ATNF): \url{https://atoa.atnf.csiro.au}} ATCA observations, including observing dates, project numbers, observation frequency, time spent on source and  array configurations are presented in Table~\ref{tab:summary_obs}. All observations were carried out in `snap-shot' mode, with 1-hour of integration over a 12-hours period as a minimum. All  used the Compact Array Broadband Backend with 2048\,MHz bandwidth, centred at wavelengths of 6\,cm ($\nu$~=~4500--6500\,MHz; the midpoint at 5500\,MHz) and 13\,cm ($\nu$~=~1972 and 2150\,MHz). 

\begin{figure*}[hbt!]
\centering
 \includegraphics[width=0.9\textwidth,clip]{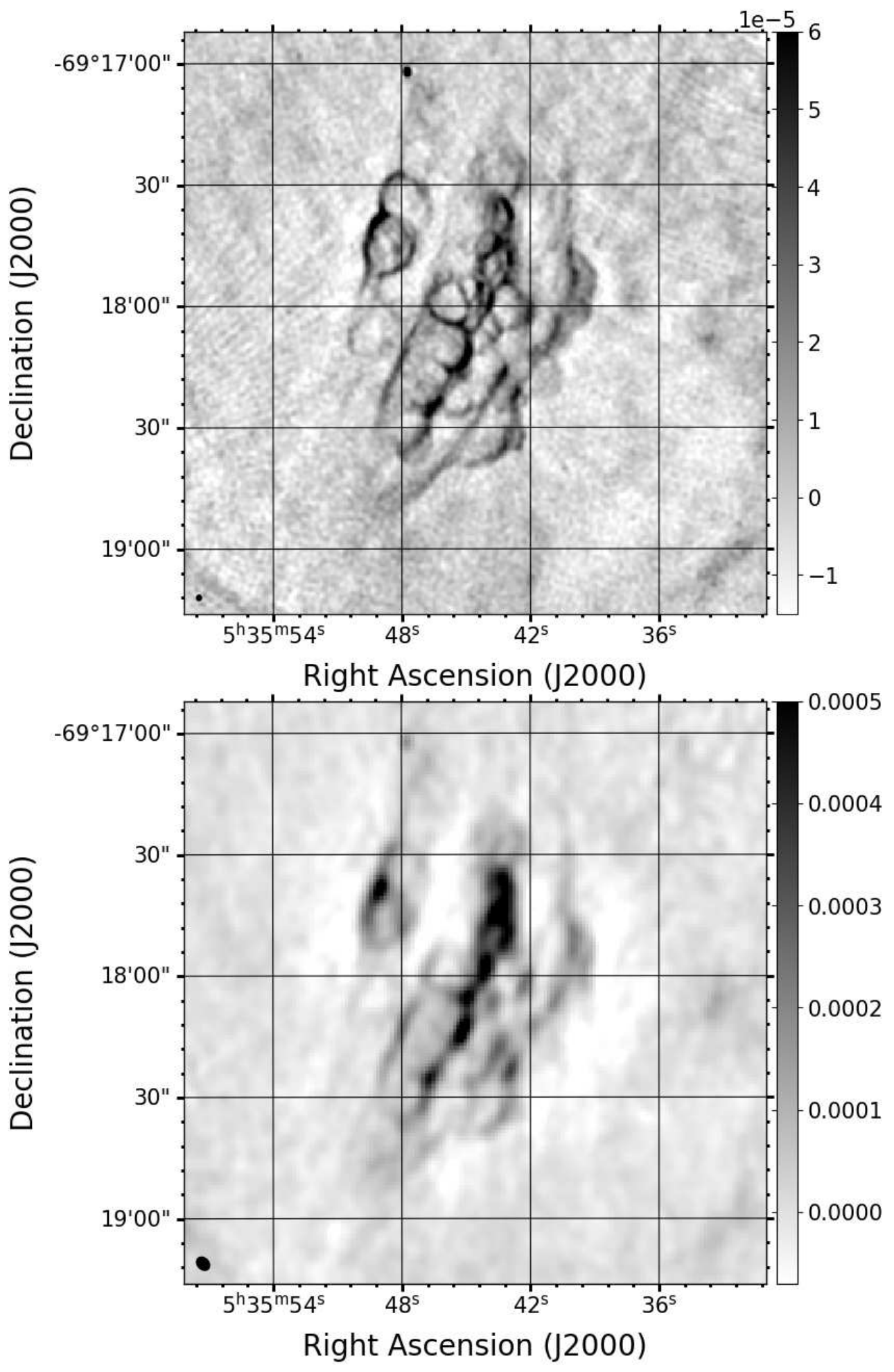}
      \caption{ ATCA intensity images of the Honeycomb Nebula at 5500\,MHz (top) and 2000\,MHz (bottom). The ellipses at the bottom left corner represent a synthesised beam of 1.3$\times$1.2\,arcsec$^2$ and 3.6$\times$2.8\,arcsec$^2$ for 5500 and 2000\,MHz, respectively. The colour bars on the right-hand side represent intensity gradients  in Jy\,beam$^{-1}$.  }
    \label{fig:1}
\end{figure*}

\begin{figure*}[hbt!]
    \centering
    \includegraphics[width=0.49\textwidth,trim=130 45 0
    0,clip]{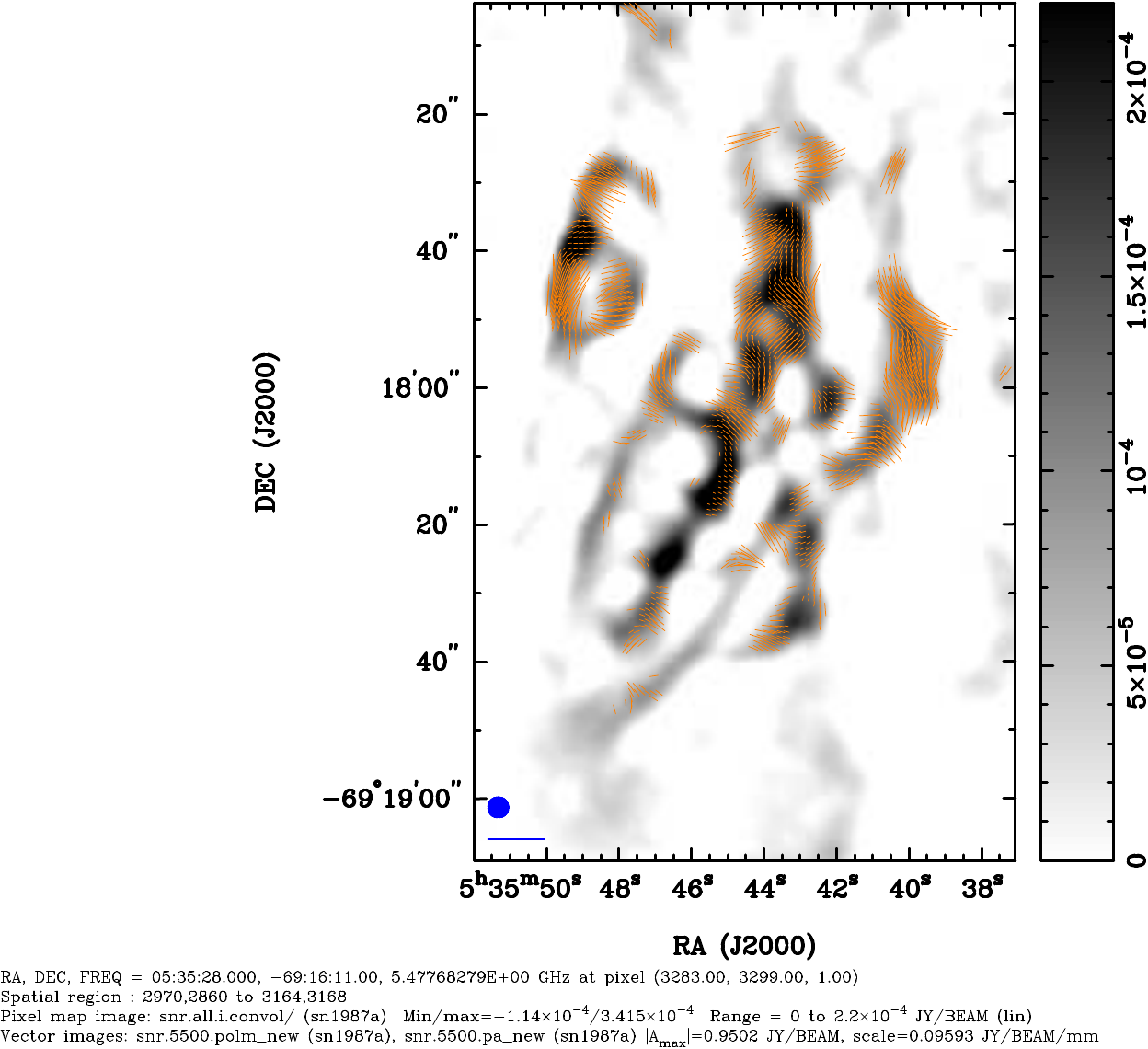}
    \includegraphics[width=0.49\textwidth,trim=0 0 0 0,clip]{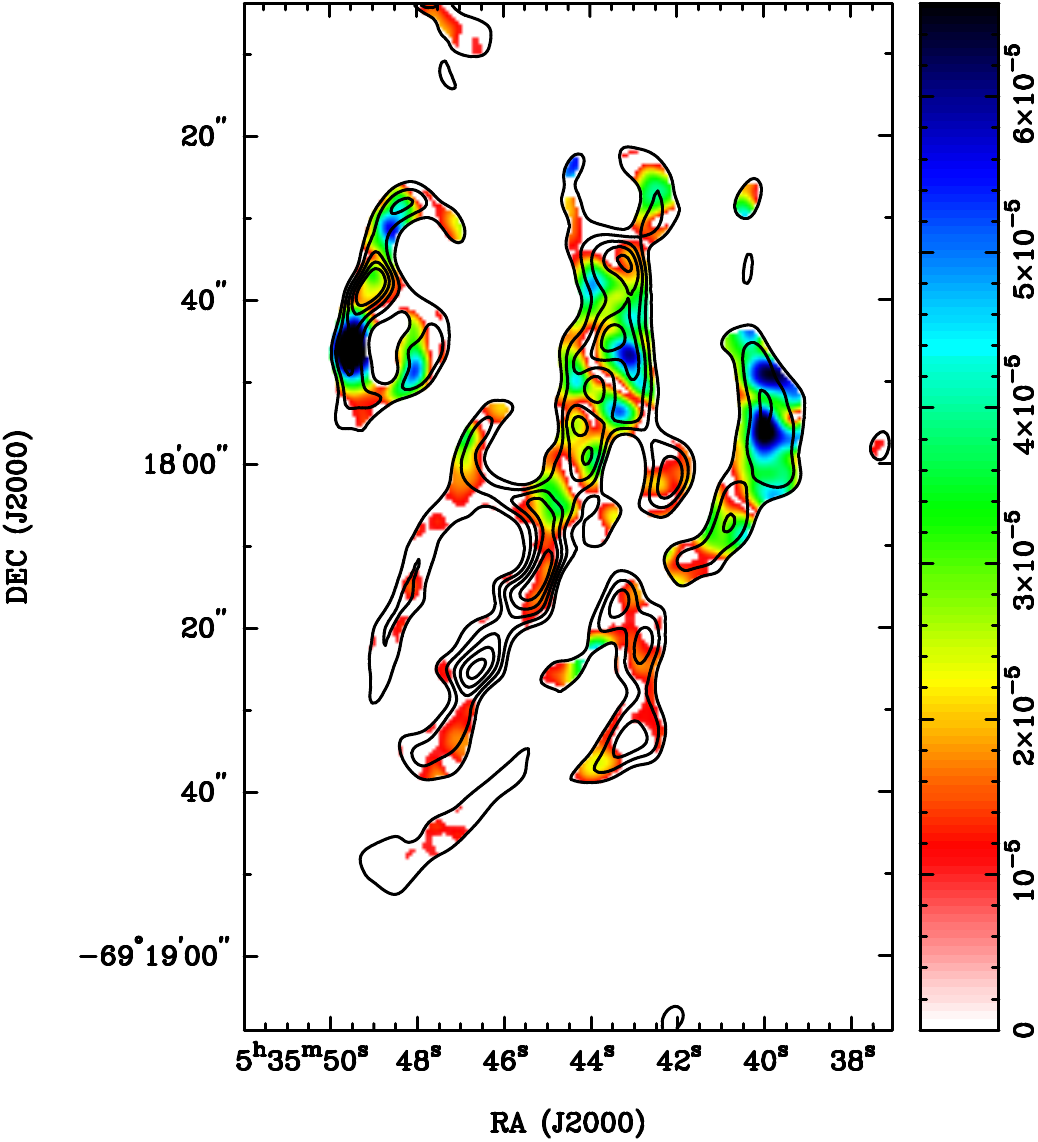}\\
    \caption{Fractional Polarisation vectors of the Honeycomb Nebula overlaid on the ATCA intensity image at 5500\,MHz (left). The blue circle at the lower left corner represents a synthesised beam of $5\times5$\,arcsec and the blue line below  represents 100\,per\,cent polarisation. The bar on the right side represent grey scale intensity gradients in Jy\,beam$^{-1}$. Polarisation intensity map of the Honeycomb Nebula at 5500\,MHz (right) with intensity  contour lines overlaid. The contour levels are 0.00005, 0.0001, 0.00015, 0.0002, and 0.00025\,Jy\,beam$^{-1}$. The colour bar on the right  represent gradients of polarisation intensity.}
    \label{fig:pol}
\end{figure*}

We used \textsc{miriad}\footnote{\url{http://www.atnf.csiro.au/computing/software/miriad/}} \citep{1995ASPC...77..433S}, \textsc{karma}\footnote{\url{http://www.atnf.csiro.au/computing/software/karma/}} \citep{1995ASPC...77..144G}, and \textsc{DS9}\footnote{\url{https://sites.google.com/cfa.harvard.edu/saoimageds9}} \citep{2003ASPC..295..489J} software packages for reduction and analysis. All observations were calibrated using the phase and flux calibrators listed in Table~\ref{tab:summary_obs} with three rounds of phase-only self-calibration using the \textsc {selfcal} task. Imaging was completed using the multi-frequency synthesis \textsc{invert} task with uniform Briggs weighting (robust~=~0) and a beam size of $1.3\times1.2$\,arcsec$^2$ for 5500\,MHz (see Fig.~\ref{fig:1}-top). We combined 2150 and 1972\,MHz data to produce one image at 2000\,MHz with a beam size of $3.6\times2.8$\,arcsec$^2$ (see Fig.~\ref{fig:1}-bottom). The \textsc {mfclean} and \textsc {restor} algorithms were used to deconvolve the images, with primary beam correction applied using the \textsc{linmos} task. We follow the same process with stokes {\it Q} and {\it U} parameters to produce polarisation maps  with a beam size of $3\times3$\,arcsec$^2$ (see Fig.~\ref{fig:pol} and Section~\ref{pol} for more details).

\begin{table*}[ht!]
	\centering
	\tiny
	\caption{A summary of  ATCA observations used in this study.}
     \vskip.25cm
	\label{tab:summary_obs}
	\begin{tabular}{@{}llccllllcc@{}}
		\hline
             Date            &Project& Array            & No.            & Bandwidth      &  Frequency     & Phase           & Flux              &Integrated time \\
          ~                  &Code   & Configuration    &   Channels     &  (MHz)         &    $\nu$ (MHz) &  Calibrator     & Calibrator        &(minutes)\\
		\hline
	   10~Apr~2010    &C015   &6A                &2048            & 2048          &2150	           &PKS\,B0530--727  & PKS\,B1934--638   & 103.2 \\
	   10~Apr~2010    &C015   &6A                &2048            & 2048          &5500	           &PKS\,B0530--727  & PKS\,B1934--638   & 187.8 \\
	   11~Apr~2010    &C015   &6A                &2048            & 2048          &5500	           &PKS\,B0530--727  & PKS\,B1934--638   & 491.4 \\ 
	   27~Jun~2010    &C015   &6C                &2048            & 2048          &1972,2150	           &PKS\,B0530--727  & PKS\,B1934--638   & 73.8 \\
	   27~Jun~2010    &C015   &6C                &2048            & 2048          &5500	           &PKS\,B0530--727  & PKS\,B1934--638   & 118.2 \\   
	   10~Jul~2010    &C015   &1.5C              &2048            & 2048          &5500  	       &PKS\,B0530--727  & PKS\,B1934--638   & 118.2 \\ 
	   19~Nov~2010    &C015   &6A                &2048            & 2048          &2150  	       &PKS\,B0530--727  & PKS\,B1934--638   & 108\\  
	   19~Nov~2010    &C015   &6A                &2048            & 2048          &5500  	       &PKS\,B0530--727  & PKS\,B1934--638   & 105 \\  
	   22~Jan~2011    &C015   &6A                &2048            & 2048          &2150  	       &PKS\,B0530--727  & PKS\,B1934--638   & 162.6  \\
	   22~Jan~2011    &C015   &6A                &2048            & 2048          &5500  	       &PKS\,B0530--727  & PKS\,B1934--638   & 162.6  \\
	   25~Jan~2011    &C015   &6A                &2048            & 2048          &5500	           &PKS\,B0530--727  & PKS\,B1934--638   & 475.2  \\
	   20~Mar~2011    &C015   &1.5A              &2048            & 2048          &2150  	       &PKS\,B0530--727  & PKS\,B1934--638   & 132.6\\
	   20~Mar~2011    &C015   &1.5A              &2048            & 2048          &5500  	       &PKS\,B0530--727  & PKS\,B1934--638   & 123.6  \\
	   22~Apr~2011    &C015   &6A                &2048            & 2048          &5500  	       &PKS\,B0530--727  & PKS\,B1934--638   & 469.8  \\
	   23~Apr~2011    &C015   &6A              &2048            & 2048          &1972,2150  	       &PKS\,B0530--727  & PKS\,B1934--638   & 133.2 \\
	   23~Apr~2011    &C015   &6A              &2048            & 2048          &5500  	       &PKS\,B0530--727  & PKS\,B1934--638   & 133.2  \\
	   27~Aug~2011    &C015   &6B                &2048            & 2048          &1972 	       &PKS\,B0530--727  & PKS\,B1934--638   & 147.6  \\
	   27~Aug~2011    &C015   &6B                &2048            & 2048          &5500 	       &PKS\,B0530--727  & PKS\,B1934--638   & 148.2  \\
	   20~Nov~2011    &C015   &1.5D              &2048            & 2048          &1972,2150 	       &PKS\,B0530--727  & PKS\,B1934--638   & 147.6  \\
	   20~Nov~2011    &C015   &1.5D              &2048            & 2048          &5500 	       &PKS\,B0530--727  & PKS\,B1934--638   & 135.6  \\
	   12~Jan~2012    &C015   &6A                &2048            & 2048          &5500 	       &PKS\,B0530--727  & PKS\,B1934--638   & 450  \\
	   16~Feb~2012    &C015   &6A                &2048            & 2048          &1972,2150 	       &PKS\,B0530--727  & PKS\,B1934--638   & 148.2 \\
	   16~Feb~2012    &C015   &6A                &2048            & 2048          &5500 	       &PKS\,B0530--727  & PKS\,B1934--638   & 133.2 \\
	   11~Apr~2012    &C015   &1.5B              &2048            & 2048          &1972,2150  	       &PKS\,B0530--727  & PKS\,B1934--638   & 118.2  \\
	   11~Apr~2012    &C015   &1.5B              &2048            & 2048          &5500  	       &PKS\,B0530--727  & PKS\,B1934--638   & 118.2  \\
	   06~Jun~2012    &C015   &6D                &2048            & 2048          &1972,2150  	       &PKS\,B0530--727  & PKS\,B1934--638   & 186  \\
	   06~Jun~2012    &C015   &6D                &2048            & 2048          &5500  	       &PKS\,B0530--727  & PKS\,B1934--638   & 133.8  \\
	   01~Sep~2012    &C015   &6A                &2048            & 2048          &5500 	       &PKS\,B0530--727  & PKS\,B1934--638   & 497.4  \\
	   02~Sep~2012    &C015   &6D                &2048            & 2048          &1972,2150 	       &PKS\,B0530--727  & PKS\,B1934--638   & 147.6  \\
	   02~Sep~2012    &C015   &6D                &2048            & 2048          &5500 	       &PKS\,B0530--727  & PKS\,B1934--638   & 133.2  \\
	   24~Nov~2012    &C015   &1.5C              &2048            & 2048          &1972,2150 	       &PKS\,B0530--727  & PKS\,B1934--638   & 133.2  \\
	   24~Nov~2012    &C015   &1.5C              &2048            & 2048          &5500 	       &PKS\,B0530--727  & PKS\,B1934--638   & 124.8  \\
	   09~Dec~2012    &C015   &6B                &2048            & 2048          &5500 	       &PKS\,B0530--727  & PKS\,B1934--638   & 384  \\
	   03~Jan~2013    &C015   &1.5D              &2048            & 2048          &1972,2150  	       &PKS\,B0530--727  & PKS\,B1934--638   & 200.4  \\
	   03~Jan~2013    &C015   &1.5D              &2048            & 2048          &5500  	       &PKS\,B0530--727  & PKS\,B1934--638   & 192  \\
	   07~Mar~2013    &C015   &6A                &2048            & 2048          &5500 	       &PKS\,B0530--727  & PKS\,B1934--638   & 482.4  \\
	   03~May~2013    &C015   &6C                &2048            & 2048          &1972,2150  	       &PKS\,B0530--727  & PKS\,B1934--638   & 118.8  \\
	   03~May~2013    &C015   &6C                &2048            & 2048          &5500  	       &PKS\,B0530--727  & PKS\,B1934--638   & 115.2 \\
	   04~May~2013    &C015   &6C                &2048            & 2048          &5500 	       &PKS\,B0530--727  & PKS\,B1934--638   & 488.4  \\
	   17~Jul~2013    &C015   &6A                &2048            & 2048          &5500  	       &PKS\,B0530--727  & PKS\,B1934--638   &  553.2  \\
	   19~Jul~2013    &C015   &6A                &2048            & 2048          &1972,2150 	       &PKS\,B0530--727  & PKS\,B1934--638   & 118.2  \\
	   19~Jul~2013    &C015   &6A                &2048            & 2048          &5500 	       &PKS\,B0530--727  & PKS\,B1934--638   & 117  \\
	   30~Aug~2013    &C015   &1.5A              &2048            & 2048          &1972,2150	           &PKS\,B0530--727  & PKS\,B1934--638   & 207  \\
	   30~Aug~2013    &C015   &1.5A              &2048            & 2048          &5500	           &PKS\,B0530--727  & PKS\,B1934--638   & 207.6  \\
	   09~Nov~2013    &C015   &6A                &2048            & 2048          &5500 	       &PKS\,B0530--727  & PKS\,B1934--638   & 512.4  \\
	   16~Apr~2014    &C015   &6A                &2048            & 2048          &5500 	       &PKS\,B0530--727  & PKS\,B1934--638   &  493.2  \\
	   27~Aug~2014    &C015   &6B                &2048            & 2048          &5500  	       &PKS\,B0530--727  & PKS\,B1934--638   & 62.4  \\
	   18~Oct~2014    &C015   &1.5A              &2048            & 2048          &5500 	       &PKS\,B0530--727  & PKS\,B1934--638   &  453  \\
	   19~Oct~2014    &C015   &1.5A              &2048            & 2048          &2150 	       &PKS\,B0530--727  & PKS\,B1934--638   & 133.2  \\
	   19~Oct~2014    &C015   &1.5A              &2048            & 2048          &5500 	       &PKS\,B0530--727  & PKS\,B1934--638   & 129.6  \\
	   02~May~2015    &C015   &6A                &2048            & 2048          &2150 	       &PKS\,B0530--727  & PKS\,B1934--638   & 148.2  \\
	   02~May~2015    &C015   &6A                &2048            & 2048          &5500 	       &PKS\,B0530--727  & PKS\,B1934--638   &  171.6  \\
	   05~Dec~2015    &C015   &1.5A              &2048            & 2048          &2150 	       &PKS\,B0530--727  & PKS\,B1934--638   & 133.2  \\
	   05~Dec~2015    &C015   &1.5A              &2048            & 2048          &5500 	       &PKS\,B0530--727  & PKS\,B1934--638   & 133.2  \\
	   08~Mar~2016    &C015   &6B                &2048            & 2048          &2150 	       &PKS\,B0530--727  & PKS\,B1934--638   & 148.2  \\
	   08~Mar~2016    &C015   &6B                &2048            & 2048          &5500 	       &PKS\,B0530--727  & PKS\,B1934--638   & 132.6  \\
	   04~Jun~2016    &C015   &1.5B              &2048            & 2048          &2150 	       &PKS\,B0530--727  & PKS\,B1934--638   & 143.4  \\
	   04~Jun~2016    &C015   &1.5B              &2048            & 2048          &5500 	       &PKS\,B0530--727  & PKS\,B1934--638   & 133.2  \\
	   24~Aug~2016    &C015   &6C                &2048            & 2048          &2150	    	   &PKS\,B0530--727  & PKS\,B1934--638   & 133.2  \\
	   24~Aug~2016    &C015   &6C                &2048            & 2048          &5500	    	   &PKS\,B0530--727  & PKS\,B1934--638   & 133.2  \\
	   22~Nov~2016    &C015   &6A                &2048            & 2048          &2150  	       &PKS\,B0530--727  & PKS\,B1934--638   & 133.2  \\
	   22~Nov~2016    &C015   &6A                &2048            & 2048          &5500  	       &PKS\,B0530--727  & PKS\,B1934--638   & 123.6  \\
	   07~Feb~2017    &C015   &6D                &2048            & 2048          &5500 	       &PKS\,B0530--727  & PKS\,B1934--638   & 415.8  \\
	   08~Feb~2017    &C015   &6D                &2048            & 2048          &2150 	       &PKS\,B0530--727  & PKS\,B1934--638   & 192  \\
	   08~Feb~2017    &C015   &6D                &2048            & 2048          &5500 	       &PKS\,B0530--727  & PKS\,B1934--638   & 177.6  \\
	   18~Feb~2017    &C015   &6D                &2048            & 2048          &2150  	       &PKS\,B0530--727  & PKS\,B1934--638   & 133.2  \\
	   18~Feb~2017    &C015   &6D                &2048            & 2048          &5500  	       &PKS\,B0530--727  & PKS\,B1934--638   & 131.4  \\
	   29~Aug~2017    &C015   &1.5A              &2048            & 2048          &5500  	       &PKS\,B0530--727  & PKS\,B1934--638   & 249  \\
	   10~Nov~2017    &C015   &1.5C              &2048            & 2048          &5500  	       &PKS\,B0530--727  & PKS\,B1934--638   & 118.2  \\
	   11~Nov~2017    &C015   &1.5C              &2048            & 2048          &5500 	       &PKS\,B0530--727  & PKS\,B1934--638   & 325.8  \\
	   30~Dec~2017    &C015   &6C                &2048            & 2048          &5500 	       &PKS\,B0530--727  & PKS\,B1934--638   & 508.2  \\
	   
		\hline
	\end{tabular}
\end{table*}

\begin{table*}[ht!]
\caption{Flux density measurements of the Honeycomb Nebula.}
\vskip.25cm
\centering
\begin{tabular}{@{}cclcc@{}}
\hline\hline
Frequency &    Flux &  Flux error  & Telescope & Reference\\
(MHz)     &     (Jy)  &       (Jy)     \\
 \hline%
888       &    0.181   &    0.0181       &ASKAP      &  \cite{2021MNRAS.506.3540P}\\
2000      &    0.110   &    0.0220       &ATCA       &  This work\\
5500      &    0.046   &    0.0089       &ATCA       &  This work\\
\hline\hline
\end{tabular}
\label{tab2}
\end{table*}

\subsection{\HI\ observations}
\begin{figure*}[hbt!]
\centering
    \includegraphics[width=\textwidth,clip]{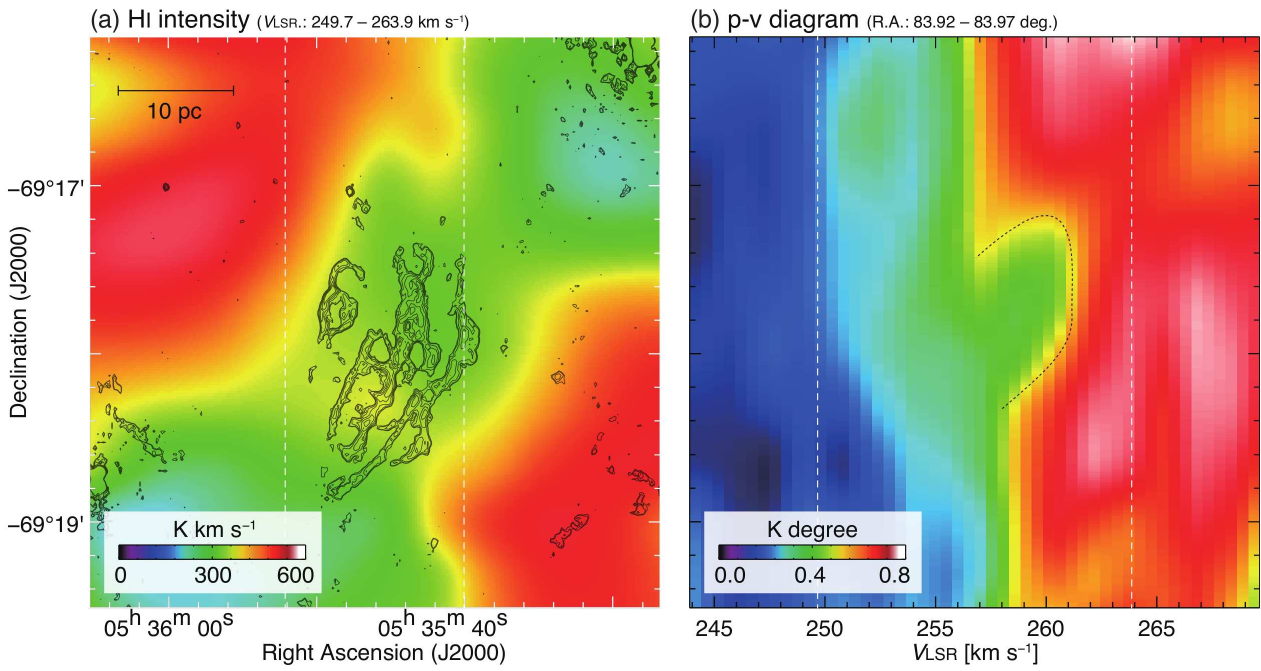}	
    \caption{(a) Velocity integrated Parkes \HI\ intensity map from 249.7 to 263.9 km s$^{-1}$ superposed with  ATCA 5500~MHz  intensity contours.  Contour levels are 1.0, 1.2, 1.8, 2.8, 4.2, and $6.0 \times 10^{-5}$ Jy beam$^{-1}$. (b) \HI\ position--velocity (p-v) diagram having an integration range from RA 83.92 to 83.97~degrees. The dashed curve indicates the boundary of a \HI\ hollowed-out structure.}
    \label{fig:HI}
\end{figure*}

In order to better understand the interstellar environment surrounding the Honeycomb Nebula, we used  archived \HI\ data from the  ATCA \&  Parkes 64-m radio telescope \citep{2003ApJS..148..473K}. The combined \HI\ data has an angular resolution of $\sim$$60''$, corresponding to a spatial resolution of $\sim$15~pc. Typical noise ﬂuctuations are $\sim$0.3~K at a velocity resolution of $\sim$1.6~km~s$^{-1}$ (Fig.~\ref{fig:HI}).

\subsection{X-ray observations}
\label{xray}
\indent
Detailed comparisons of  X-ray morphology with our new radio images were made using sub-arcsecond resolution data from the \textit{Chandra} X-ray Observatory \citep{Weiss1996}. Due to its proximity to SN~1987A, the Honeycomb Nebula has been observed many times by  \textit{Chandra}. We retrieved all available data that included the Honeycomb Nebula  using the Advanced CCD Imaging Spectrometer S-array \cite[ACIS-S,][]{Garmire2003}  with exposure times of $\gtrsim$40~ks. A list of observations are given in Table~\ref{tab:x-ray_obs}.

We processed each of these datasets using the CIAO~v4.15\footnote{See \url{http://cxc.harvard.edu/ciao/}} \citep{Fru2006} software package with CALDB~v4.10.7\footnote{See \url{http://cxc.harvard.edu/caldb/}}. The data were reduced using the contributed script \texttt{chandra\_repro}, resulting in the filtered exposure times listed in Table~\ref{tab:x-ray_obs}. 

\begin{table}
\caption{Details of \textit{Chandra} X-ray observations.}
\begin{center}
\label{tab:x-ray_obs}
{\renewcommand{\arraystretch}{1.3}
\begin{tabular}{cccc}
\hline
\hline
Obs. ID & Date & PI & Exp. time \\
 & & & (ks) \\
\hline
1967 & 2000-12-07 & R. McCray & 98.76 \\
2831 & 2001-12-12 & D. Burrows & 49.41 \\
2832 & 2002-05-15 & D. Burrows & 44.26 \\
3829 & 2002-12-31 & D. Burrows & 49.01 \\
3830 & 2003-07-08 & D. Burrows & 45.31 \\
4614 & 2004-01-02 & D. Burrows & 46.49 \\
4615 & 2004-07-22 & D. Burrows & 48.83 \\
\hline
\end{tabular}}
\end{center}
\end{table}%

\begin{figure*}[hbt!]
\centering
 \includegraphics[width=0.9\textwidth,clip]{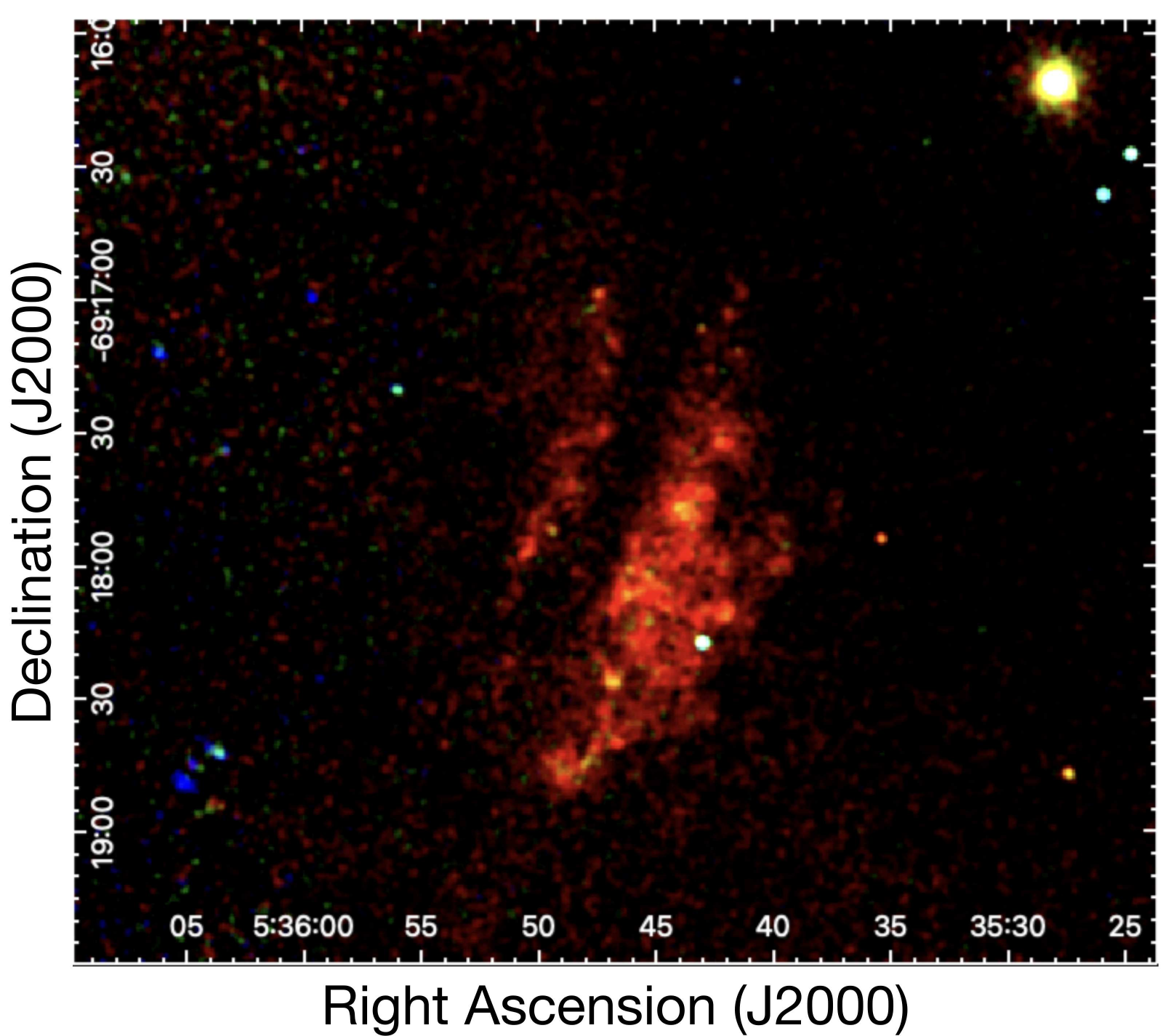}
    \caption{Three-colour \textit{Chandra} image of the Honeycomb Nebula. RGB colours are 0.5--1~keV, 1--2~keV, and 2--8~keV. The image has been smoothed using a 3$\times$3 pixel Gaussian kernel. SN~1987A is the bright source in the top-right of the image.}
    \label{fig:x-ray_im}
\end{figure*}

We reprojected the resulting level 2 event files from each observation to a common tangent point using  CIAO  \texttt{reproject\_obs}  before merging them using  \texttt{merge\_obs}  for a combined exposure time of $\sim382$~ks. Finally, flux images were produced from the merged event file at 0.5--1~keV, 1--2~keV and 2--8~keV  using  \texttt{fluximage} . The resulting three-colour image is shown in Fig.~\ref{fig:x-ray_im}.

\subsection{Optical observations}
\indent

\begin{figure*}[hbt!]
\centering
		\includegraphics[width=\textwidth,clip]{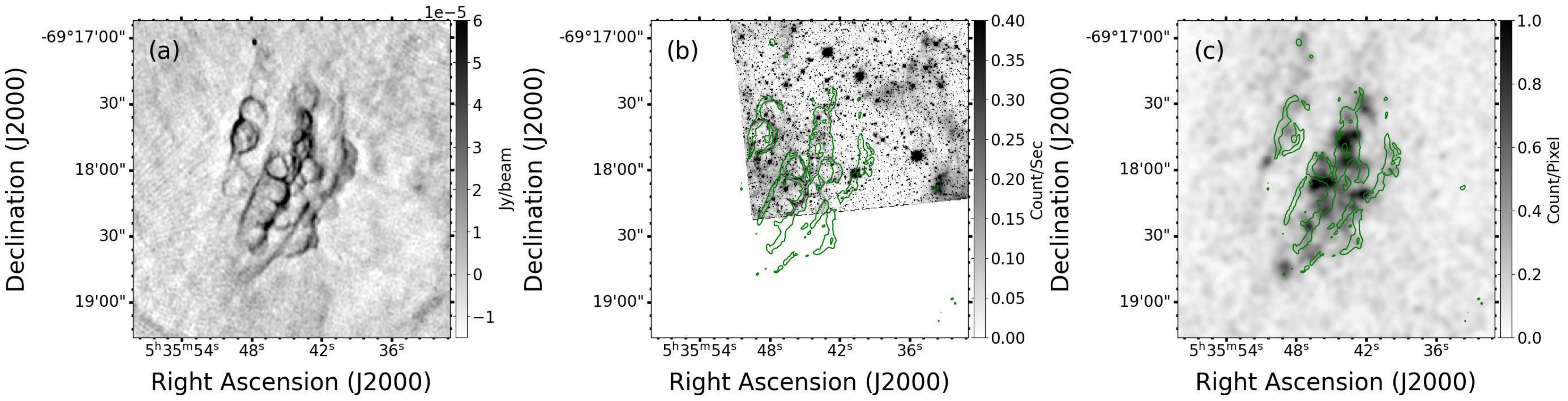}
    \caption{ Honeycomb Nebula: a) our new ATCA image at 5500\,MHz, b) partial HST image, and c) smoothed \textit{Chandra} (0.5--1\,keV) image. Green contours are from the 5500\,MHz image at 20\,$\mu$Jy. }
    \label{fig:2}
\end{figure*}

\begin{figure*}
\centering
		\includegraphics[width=\textwidth,clip]{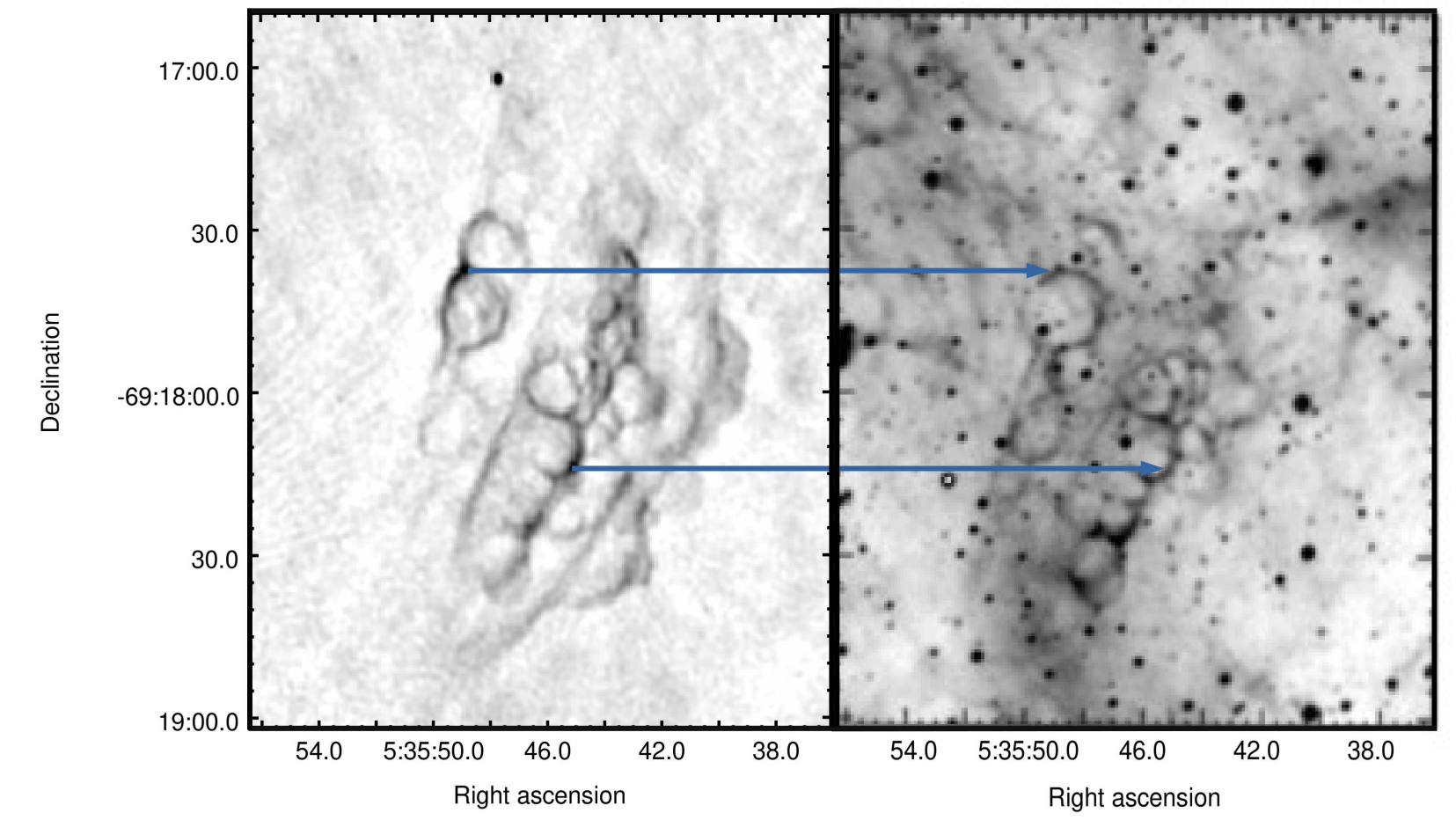}
    \caption{ATCA image of the Honeycomb Nebula at 5500\,MHz (left) with  corresponding regions in the \Halpha+\NII\ image (right) from  \citet{2010MNRAS.408.1249M}.  Note the west and south-west regions of the remnant seem to disappear in the \Halpha+\NII\ image. }
    \label{fig:3}
\end{figure*}

HST images of the Honeycomb Nebula were downloaded from the Mikulski Archive for Space Telescopes\footnote{\url{https://mast.stsci.edu/portal/Mashup/Clients/Mast/Portal.html}}. We used \textsc{swarb}\footnote{\url{https://www.astromatic.net/software/swarp/}} \citep{2002ASPC..281..228B} to stitch four images together to produce a single image (Fig.~\ref{fig:2}). We also use the \Halpha+\NII\ image of the Honeycomb Nebula from \citet{2010MNRAS.408.1249M} (Fig.~\ref{fig:3}) for comparison.

\section{RESULTS AND DISCUSSION} 
 \label{res}

\subsection{Radio morphology}
 \indent
 \label{mor}


In Fig.~\ref{fig:1}, we present deep high-resolution radio-continuum images of the Honeycomb Nebula at 5500 and 2000\,MHz. 
Its radio emission is primarily non-thermal, based on a steep spectral index (see Section~\ref{SI}). 
The east and south-east parts of the remnant appears as if a blast wave expanded into a rarefied environment and then collided with nearby dense gas \citep{1993MNRAS.263L...6M}. Its mid-section is bright, as especially noted in the 2000\,MHz image (Fig.~\ref{fig:1}-bottom). Using our images, we calculated the centre of the Honeycomb Nebula at RA~(J2000)~=~5$^{h}$35$^{m}$45.16$^{s}$,  Dec~(J2000)~=~--69$^\circ$17$'$59$''$95 with a size of $\sim59 \times35$\,arcsec$^2$ ($\sim14 \times8$\,pc).

To better understand the morphology of the  SNR, we compare our new 5500\,MHz ATCA image with HST and \textit{Chandra} images as shown in Fig.~\ref{fig:2}. We note  the radio emission 
follows the optical emission somewhat (see Fig.~\ref{fig:2} b), but there is no correlation with X-ray emission, which we believe it is  predominantly  thermal  (Fig.~\ref{fig:2} c). The west and south-west side of the remnant disappears in the \Halpha+\NII\ image (Fig.~\ref{fig:3}) which support  heterogeneous densities as noted in the blast wave propagation scenario stated above. 

Perhaps  most similar  to the  morphology of the Honeycomb Nebula is MCSNR~J0052--7236 (DEM\,S68) in the Small Magellanic Cloud, which is significantly larger with a diameter of 88 \,pc.  Additional information can be found in \citet{2005MNRAS.364..217F,2007MNRAS.376.1793P,2008A&A...485...63F,2011A&A...530A.132O,2012A&A...545A.128H,2014AJ....148...99C,2015ApJ...803..106R,2019MNRAS.486.2507A,2019MNRAS.485L...6G,2019A&A...631A.127M,2019MNRAS.490.1202J}  and Cotton et al. (in prep.).

\subsection{Spectral index}
 \indent
 \label{SI}

\begin{figure}[hbt!]
    \centering
    \includegraphics[width=\columnwidth,trim=0 0 0 0,clip]{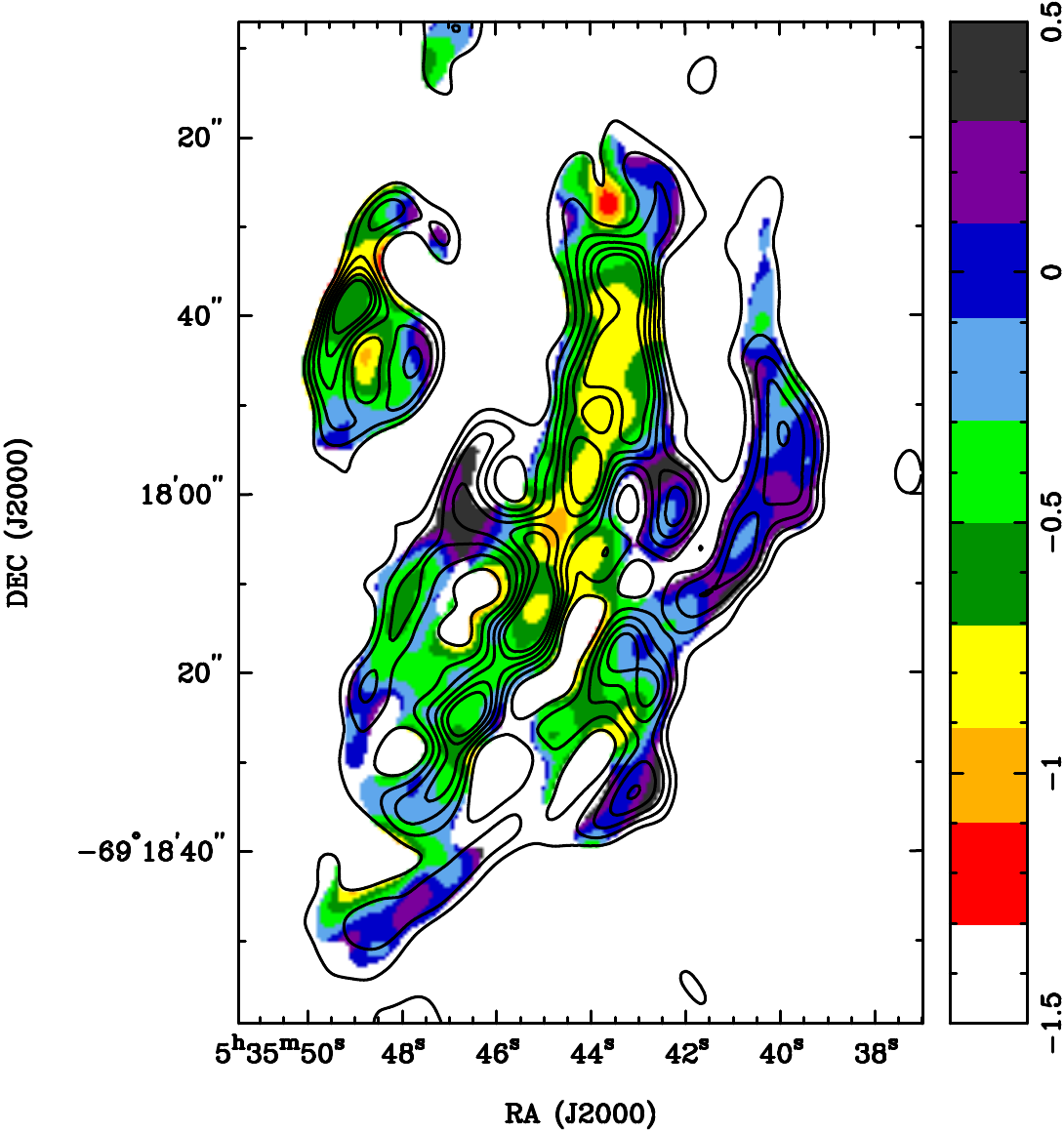}
    \caption{Spectral index map of the Honeycomb Nebula (ATCA; 2000, 5500\,MHz) with 5500\,MHz contour lines overlaid. The contour levels are 0.00005, 0.0001, 0.00015, 0.0002,  0.00025 and 0.0003\,Jy\,beam$^{-1}$. The colour bar on the right-hand side represent gradients of  spectral index.}
    \label{fig:4}
\end{figure}

The radio spectral index $\alpha$ can be defined as the slope in power-law dependence of flux density $S_{\nu}$ on frequency $\nu$: $S_{\nu}$~$\propto$~$\nu^\alpha$. We produced a spectral index map for the Honeycomb Nebula with the 2000 and 5500\,MHz images (Fig.~\ref{fig:4}) by first re-gridding the images to the finest  pixel size ($0.4\times0.4$\,arcsec$^2$) using the \textsc{miriad} task \textsc{regrid}. These were then smoothed to a common resolution ($4\times4$\,arcsec$^2$) using  \textsc{convol}, after which the task \textsc{maths} created the spectral index map (Fig.~\ref{fig:4}) in a similar way as LMC MCSNR~0624--6948 \citep{2022MNRAS.512..265F}. The spectral index values for the west and south-west sides of the remnant are changing from --0.5 to 0.5, while the east, north-east and main body show steep values ($<$--0.5). Interestingly, we notice a very steep region (--1.5) towards the north (see Fig.~\ref{fig:4}).   

\begin{figure}[hbt!]
    \centering
    \includegraphics[width=\columnwidth,trim=0 0 0 0,scale=1.5,clip]{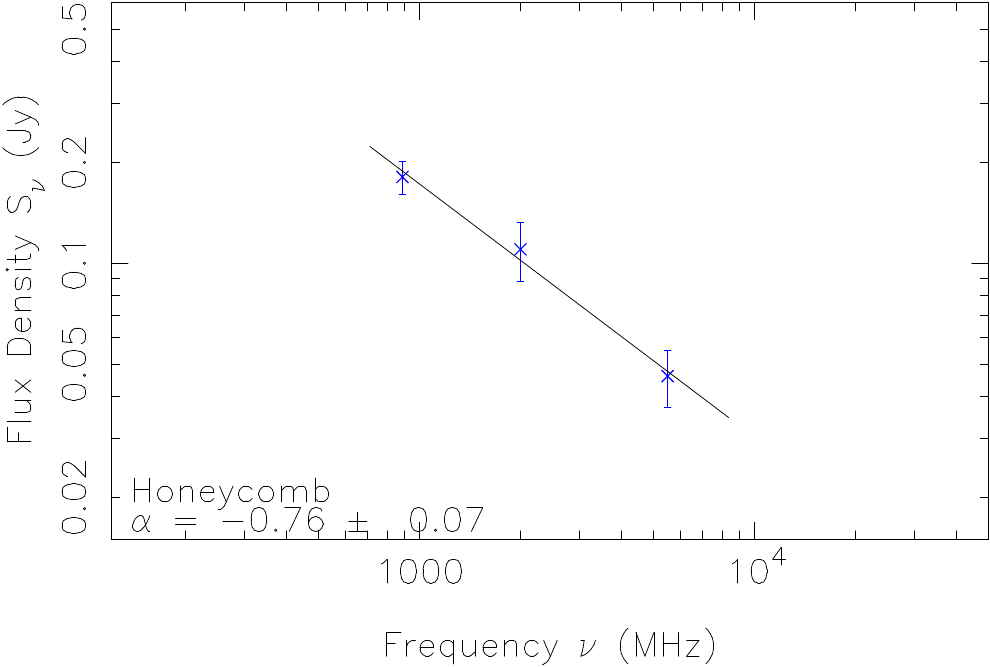}
    \caption{Radio continuum spectrum of the Honeycomb Nebula. }
    \label{fig:5}
\end{figure}

In Fig.~\ref{fig:5}, we plot integrated flux densities at three frequencies in order to estimate the Honeycomb Nebula's overall spectral index. Our observations were combined with flux density measurements from  ASKAP \citep{2021MNRAS.506.3540P} (see Table~\ref{tab2}). We acknowledge errors are introduced when measuring ATCA flux densities because of missing short spacing inherent in the data. 
In the figure, the black line represents the best power-law weighted least-squares fit wherein we find a spatially integrated spectral index of $\langle\alpha\rangle= -0.76 \pm 0.07$, flatter than a previous estimation of --1.2  \citep{1995AJ....109.1729C}.  This may be indicative of \textbf{a} younger aged SNR as our value is steeper than  average shell-type SNRs, as observed for the Galaxy and a range of nearby galaxies, including the LMC \citep{2012SSRv..166..231R,2014Ap&SS.354..541U,2014SerAJ.189...15G,2019A&A...631A.127M,book2,2017ApJS..230....2B,2023MNRAS.518.2574B}. Although there are regions with $\alpha > 0$, this strongly indicates that non-thermal radio emission dominates across the SNR.

%
 
\subsection{Polarisation}
 \indent
 \label{pol}

Fractional polarisation ($P$) can be calculated using the equation:
\begin{equation}
\label{eq}
P=\frac{\sqrt{S^2_Q+S^2_U}}{S_I},
\end{equation}

\noindent where $P$ is the mean fractional polarisation, $S_{Q}$, $S_{U}$, and $S_{I}$ are intensities for the $Q$, $U$, and $I$ Stokes parameters, respectively. 

We used the \textsc{miriad} task \textsc{impol} to produce polarisation maps for the Honeycomb Nebula. We find the fractional polarisation is generally in the radial direction and prominent in the east, west, and midsection of the remnant (Fig.~\ref{fig:pol}). Its average $P$ value at 5500\,MHz is $24\pm5$\% with a maximum value of $95\pm16$\%.

\subsection{Equipartition model} 
There are three standard ways for determination of magnetic field strengths in the ISM: from Zeeman splitting, rotation measure, and by using so-called equipartition calculation (eqp). Zeeman splitting method is appropriate only for very dense phases of ISM, like cores of molecular clouds. Rotation measure method is based on polarisation of radio emission from radio sources. The eqp method is based on approximately equpartition, or constant partition between the cosmic rays (CRs) and magnetic field energy densities. 

To estimate the magnetic field, we used eqp model\footnote{\url{http://poincare.matf.bg.ac.rs/~arbo/eqp}}. This method uses modeling and simple parameters to estimate intrinsic magnetic field strength and energy contained in the magnetic field and cosmic ray particles using radio synchrotron emission \citep{2012ApJ...746...79A,2013ApJ...777...31A,2018ApJ...855...59U}.

This approach is a purely analytical, described as a rough, only order of magnitude estimate because of assumptions used in analytical derivations, and errors in determination of distance, angular diameter, spectral index, a filling factor, and flux density, tailored especially for the  magnetic field strength in SNRs. \citet{2012ApJ...746...79A,2013ApJ...777...31A,2018ApJ...855...59U} present two models; the difference is in assumption whether there is eqp, precisely constant partition with CRs or only CR electrons. \citet{2018ApJ...855...59U} showed the latter type of eqp is better assumption to the former.


Using the \citet{2018ApJ...855...59U} model\footnote{We use: $\alpha=0.76$, $\theta$~=~1.22\,arcmin, $\kappa=0$, $S_{\rm 5500\,MHz}$~=~0.046\,Jy, and f~=~0.25.}, the mean eqp field over the whole Honeycomb Nebula is $48\pm5$\,$\mu$G, with an estimated minimum energy of $E_{\rm min}=3\times10^{49}$\,erg. As this SNR is young and expanding into a low density environment (see Section~\ref{Surface}), the magnetic field has to be amplified, not only compressed by shock wave. (The original model, cited in \citet{2012ApJ...746...79A}, yields a mean eqp field of $80\pm5$\,$\mu$G, with an estimated minimum energy of $E_{\rm min}=9\times10^{49}$\,erg.)

\subsection{\HI\ distribution}
 \indent
 \label{HI}

Fig.~\ref{fig:HI}(a) shows the integrated intensity map of \HI\ towards the Honeycomb Nebula. There is an intensity gradient of \HI\ increasing from the centre of the SNR to the north-east and south-west. The brightest feature ($\sim$600~K~km s$^{-1}$) is roughly twice higher than the lowest  towards the SNR ($\sim$300~K~km s$^{-1}$). On the other hand, the north-west and south-west regions show the lowest \HI\ values ($\sim$200--300~K~km~s$^{-1}$); roughly consistent with the direction of the SNR.

Fig.~\ref{fig:HI}(b) shows the position--velocity diagram of \HI\ towards the Honeycomb Nebula. We find the \HI\ intensity is hollowed out along the dashed curve, the velocity range of which is from 257 to 262~km~s$^{-1}$. Moreover, the spatial extent of  hollowed-out \HI\ is roughly consistent with the radio size of the SNR. We argue these hollowed-out distributions of  \HI\ indicate an expanding gas motion from stellar winds originating with the progenitor system. The expanding velocity of \HI\ in the Honeycomb Nebula is estimated to be $\Delta V \sim$5 km s$^{-1}$, which is roughly simular to Magellanic SNRs RX~J0046.5$-$7308 \citep[$\Delta V \sim$3--5~km~s$^{-1}$][]{2019ApJ...881...85S} and N132D \citep[$\Delta V \sim$6~km~s$^{-1}$][]{2020ApJ...902...53S}. If our interpretation is correct, the Honeycomb Nebula exploded inside a low-density wind cavity. To test this scenario, we need further \HI\ observations with finer angular resolutions.

\subsection{X-ray emission}
 \indent

Our three-colour \text{Chandra} image of the Honeycomb Nebula is shown in Fig.~\ref{fig:x-ray_im}. As shown in Fig.~\ref{fig:2}(c), its morphology is broadly similar the radio morphology of our new ATCA images, with its brightness increasing towards the western side where the SNR is thought to be evolving into a denser ambient medium. However, as noted above, on small scales the X-ray and radio emission are not well correlated, which is expected given their thermal and non-thermal origins, respectively.

\citet{2016A&A...585A.162M} performed an X-ray spectral analysis of the Honeycomb Nebula using 
\textit{XMM-Newton} data, finding that its emission is soft with a plasma temperature ($kT$) of $\sim0.3$~keV and  faint with an X-ray luminosity ($L_{\mathrm{X}}$) of $\sim4\times10^{34}$~erg~s$^{-1}$ in the 0.3--8~keV range. In addition, the spectra showed enhanced abundances of $\alpha$-process elements, which along with the star formation history in the region led to \citet{2016A&A...585A.162M} assigning a high probability that the Honeycomb Nebula resulted from a core-collapse event. 

Given the small size of the nebula and assuming  size is a useful proxy for age, the relatively low observed luminosity and temperature  are unusual. Another example of such an SNR is MCSNR~J0512--6707 reported by \citet{2015A&A...583A.121K}, which is slightly larger than the Honeycomb Nebula though somewhat cooler and fainter. These authors suggest that MCSNR~J0512--6707 initially evolved into the wind-blown cavity of the progenitor and is now interacting with the swept-up shell. If the Honeycomb Nebula resulted from a similar interaction, as we suggest above, the low $kT$ and $L_{\mathrm{X}}$ observed  further support this interpretation.

\subsection{The Honeycomb SNR surface brightness} 
 \indent
 \label{Surface}

\begin{figure}[hbt!]
    \centering
    \includegraphics[width=\columnwidth,trim=0 0 0 0,scale=1.5,clip]{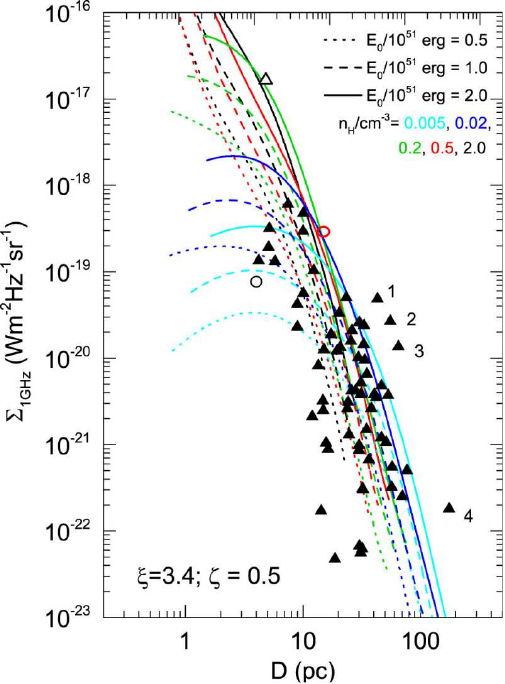}
    \caption{Radio surface brightness–to–diameter diagram for SNRs at a frequency of 1\,GHz (black triangles), obtained from numerical simulations \citep[][their Fig.~3]{2018ApJ...852...84P}. The Honeycomb Nebula is marked with an open red circle while the open triangle represents Cassiopeia\,A. The open circle represents the youngest Galactic SNR, G1.9+0.3 \citep{2020MNRAS.492.2606L}. Numbers represent the following SNRs: (1) CTB\,37A, (2) Kes\,97, (3) CTB\,37B and (4) G65.1+0.6. }
    \label{fig:8}
\end{figure}

We compared a previous $\Sigma$--D diagram \citep[][their Fig.~3]{2022PASP..134f1001U,2020NatAs...4..910U,2018ApJ...852...84P} with our values, D~=~17.7\,pc and $\Sigma_{\rm 1\,GHz} =30\times10^{-20}$\,W\,m$^{-2}$\,Hz$^{-1}$\,sr$^{-1}$, for the Honeycomb Nebula.  From this comparison, we believe that the Honeycomb Nebula is undergoing expansion within a surrounding environment characterised by a low-density of 0.02--0.2\,cm$^{-3}$ (Fig.~\ref{fig:8}). 

We should emphasise that scatter of evolutionary tracks in $\Sigma-D$ plane is obvious. It is result of SNR evolution
in different environments with different explosion energies. The
Honeycomb nebula is indeed close to the 0.2 cm$^{-3}$ and $10^{51}$ erg track, at the
brightness maximum which marks the transition between free expansion and
Sedov's phase. However, Pavlovi{\'c} et al. (2018) models are basically for SN Ia (uniform medium), and only for certain SN II models they assume a circumstellar medium with power-law density profile $~r^{-2}$. If this SNR is expanding in a cavity, then we would have approximately homogeneous density until the transition to ISM where probably there would be a dense
shell (or shells). Due to this, the evolutionary status established here should be taken with caution.

%

\section{CONCLUSIONS} 
 \indent
 \label{con}

We present new high-resolution and sensitive ATCA images for the LMC Honeycomb Nebula at 2000 and 5500\,MHz. We also compare these images with available X-ray, optical and \HI\ images.

The following is a summary of our findings:
\begin{itemize}
\item The Honeycomb Nebula has a steeper than average SNR radio spectral index of $-0.76\pm0.07$, suggesting \textbf{a} younger age,
\item A radial polarisation with an average fractional polarisation of $24\pm5$\% and maximum value of $95\pm16$\% at 5500\,MHz, also indicating \textbf{a} younger age,
\item An eqp field of $48\pm5$\,$\mu$G, also is consistent with a younger age,     
\item This SNR has the highest surface brightness, $\Sigma_{\rm 1\,GHz} =30\times10^{-20}$\,W\,m$^{-2}$\,Hz$^{-1}$\,sr$^{-1}$ for its diameter in comparison to Galactic SNRs presented in Fig. 9. It tentatively suggests previous expansion in homogeneous low density medium and after that collision with denser shell of a young SNR in the transition between late free expansion and early Sedov phase of evolution.
\end{itemize}

When combined with its unusual morphology and other data found in the literature, this all indicates that the remnant's progenitor most likely underwent a core-collapse event inside a low-density wind cavity.

%










\acknowledgements{
The Australia Telescope Compact Array (ATCA) and Australian SKA Pathﬁnder (ASKAP) are part of the Australia Telescope National Facility which is managed by CSIRO. H.S. acknowledges funding from JSPS KAKENHI Grant Number 21H01136. D.U. acknowledges the Ministry of Education, Science and Technological Development of the Republic of Serbia support through the contract No. 451-03-68/2022-14/200104, and the support through the joint project of the Serbian Academy of Sciences and Arts and Bulgarian Academy of Sciences on the detection of extragalactic SNRs and \HII\ regions.
}



\newcommand\eprint{in press }

\bibsep=0pt

\bibliographystyle{aa_url_saj}

{\small

\bibliography{sample_saj}
}



\clearpage

{\ }

\clearpage

{\ }

\newpage

\begin{strip}

{\ }



\naslov{UPU{T}{S}TVO ZA AUTORE}


\authors{R. Z. E. Alsaberi$^{1}$, M. D. Filipovi\'c$^{1}$, H. Sano$^{2,3}$, P. Kavanagh$^{4}$, P. Janas$^{4}$, J. L. Payne$^{1}$, D. Uro\v{s}evi\'c$^{5}$}

\vskip3mm


\address{$^1$Western Sydney University, Locked Bag 1797, Penrith South DC, NSW 1797, Australia}

\Email{19158264@student.westernsydney.edu.au, m.filipovic@westernsydney.edu.au, astronomer@icloud.com}

\address{$^2$Faculty of Engineering, Gifu University, 1-1 Yanagido, Gifu 501-1193, Japan}

\address{$^3$National Astronomical Observatory of Japan, Mitaka, Tokyo 181-8588, Japan}
\Email{hsano@gifu-u.ac.jp}

\address{$^4$School of Cosmic Physics, Dublin Institute for Advanced Studies, 31 Fitzwillam Place, Dublin 2, Ireland}

\Email{patrick.kavanagh@mu.ie, pawel.janas.2020@mumail.ie}

\address{$^5$Department of Astronomy, Faculty of Mathematics, University of Belgrade, Studentski trg 16, 11000 Belgrade, Serbia}
\Email{dejan.urosevic@matf.bg.ac.rs}
\vskip3mm


\centerline{{\rrm UDK} \udc}


\vskip1mm

\centerline{\rit Uredjivaqki prilog}

\vskip.7cm

\baselineskip=3.8truemm

\begin{multicols}{2}

{
\rrm

Ovaj prilog ima za cilj da pomogne autorima kod pripreme
qlanaka za na{\ss} qasopis.~Radove, po mogu{\cc}nosti, treba
pripremiti koriste{\cc}i}  \LaTeX\ {\rrm uz posebnu datoteku koja
defini{\ss}e izgled qasopisa} Serbian Astronomical Journal {\rrm i
sadr{\zz}i unapred definisana polja u preambuli dokumenta i neke
dodatne ili de\-li\-miq\-no izmenjene komande koje se koriste pri unosu
samog teksta.~Autori mogu koristiti ovaj dokument kao primer pri
kucanju svojih radova.

{\ }

}

\end{multicols}

\end{strip}


\end{document}